# Current-induced magnetization dynamics in single and double layer magnetic nanopillars grown by molecular beam epitaxy

N Müsgens[1], E Maynicke[1], M Weidenbach[1], C J P Smits[1], M Bückins[2], J Mayer[2], B Beschoten[1], and G Güntherodt[1]

[1]Physikalisches Institut IIA, RWTH Aachen University, 52056 Aachen, and Virtual Institute for Spinelectronics (VISel)

[2]Gemeinschaftslabor für Elektronenmikroskopie, RWTH Aachen University, 52056 Aachen

E-mail: bernd.beschoten@physik.rwth-aachen.de

**Abstract**

Molecular beam epitaxy is used to fabricate magnetic single and double layer junctions which are deposited in prefabricated nanostencil masks. For all Co | Cu | Co double layer junctions we observe a stable intermediate resistance state which can be reached by current starting from the parallel configuration of the respective ferromagnetic layers. The generation of spin waves is investigated at room temperature in the frequency domain by spectrum analysis, demonstrating both in-plane and out-of-plane precessions of the magnetization of the free magnetic layer. Current-induced magnetization dynamics in magnetic single layer junctions of Cu | Co | Cu has been investigated in magnetic fields which are applied perpendicular to the magnetic layer. We find a hysteretic switching in the current sweeps with resistance changes significantly larger than the anisotropic magnetoresistance effect.

1. Introduction

Spin-transfer-induced magnetization dynamics has been studied experimentally [1-7] since the theoretical predictions by Slonczewski [8] and Berger [9]. A standard spin transfer device consists of two ferromagnetic (FM) layers of unequal thicknesses, separated by a nonmagnetic layer (NM) forming a nanopillar with a lateral dimensions on the 100 nm scale. Electrons flowing perpendicular to the sample plane get spin-polarized by passing one of the ferromagnets and then repolarized at the other NM/FM interface. Thereby the spin component transverse to the second FM layer's magnetization is absorbed and acts as a torque on its magnetization. Because of the difference in layer thicknesses, only the magnetization of the thin ferromagnetic layer gets destabilized and can either be switched or excited in a precessional mode depending on the applied field and the passing current [10-12].

A sufficiently large electric current can affect the magnetization state of a ferromagnet not only in double ferromagnetic layer systems, but also in single ferromagnetic nanopillars [13-19]. A current flow generates a spin accumulation at both interfaces of the NM/FM/NM trilayer. To obtain an unequal torque on the magnetization the spin accumulations have to differ. This can be achieved by asymmetric NM leads. Again, the interface plays an important role for both switching behaviour and magnetization dynamics. One approach to better control the interfaces is the use of molecular beam epitaxy (MBE) [20]. We therefore fabricated magnetic nanopillars consisting of both magnetic single and double layer structures by MBE and analyzed their current-induced switching behaviour by magneto-transport and by high-frequency probes.

2. **Samples and experimental setup**

Focused-ion-beam (FIB) milling is used to fabricate nanostencil templates which are suited for the preparation of multilayered spin-valve structures in the current perpendicular to plane (cpp) geometry [4, 21-22]. This approach allows to quickly modify and optimize both material combinations and growth conditions. The device dimensions are defined prior to the thin film deposition. In figure 1 we summarize the relevant process steps. First, a bottom electrode is fabricated by optical lithography and subsequent $Ar^+$ etching of an extended Pt layer, which was sputtered onto a Si substrate. Subsequently, an insulator $SiO_2$ and a second Pt layer are sputtered. Next FIB (the FEI Strata 205 with $Ga^+$ liquid metal ion source has been operated at 30 keV with a beam current of 1 pA) is used to open up an aperture in the top Pt layer. The size of the aperture directly defines the diameter of the magnetic nanopillar device. The aperture in the hard mask gives access to the underlying insulator for the subsequent HF dip, which yields an isotropic selective wet etching of the $SiO_2$. The resulting undercut (for schematic picture see figure 1(a)) can easily be imaged by scanning electron microscopy in top view of the apertures (figure 1(b)). The nanostencil is then transferred into an MBE chamber for the growth of the desired thin film stack (figure 1(c)). Details about the crystallinity of the pillars will be discussed elsewhere [23]. As a final step the undercut is filled up with a thick Cu contact layer. Optical lithography and Ar$^+$ milling is then used to define a top electrode in cross-point geometry which guarantees electrical access to both top and bottom electrodes.

Transport measurements were performed at room temperature in an external magnetic field. The differential resistance *d*V/*d*I was measured in four-point geometry (see figure 2) by a lock-in technique with a 100 µA modulation current at *f* = 1132 Hz which is superimposed to a DC current. The sample is connected to a high frequency (HF) sample holder. It consists of a flexible HF cable and a coplanar waveguide (not shown in figure 2) with a total bandwidth of ~18 GHz. In order to detect the microwave emission of the junction we used a bias tee which separates DC from HF signals. The latter is amplified by 40 dB and analyzed by a 44 GHz bandwidth spectrum analyzer.

Positive current is defined by electron flow from the thin to the thick ferromagnetic layer in magnetic double layer samples while in magnetic single layer samples positive current is given by electron flow from the thin to the thick Cu metal layer.

3. **Spin transfer studies in magnetic double layer systems**

We first focus on magneto-transport and microwave emission data on magnetic double layer samples. The stack sequence of the pillar junction is | 3 nm Co | 25 nm Cu | 15 nm Co |. As an example, we discuss results on a 50 × 150 nm$^2$ junction in more detail. In figure 3 we compare the magnetic switching as obtained in the differential resistance from magnetic field sweeps (figure 3 (a)) and from current sweeps (figure 3(b)). Data are taken at room temperature with the magnetic field oriented in the sample plane close to the easy axis direction. The magneto-resistance curve was taken at a small DC current of 0.1 mA (figure 3(a)). The device shows a clear hysteretic switching between a low resistive parallel (*P*) and high resistive antiparallel (*AP*) state with a magneto-resistance value of $\Delta R/R \approx 3.1$ % (see black curve in figure 3(a)). Sweeping the magnetic field from large positive ($H > H_c$) to negative values the junction first remains in the low-resistance state for $H > 0$ and switches completely into the high-resistance state at small negative fields. Note, that in contrast to the switching at the outer coercive fields the low field switching is not abrupt. It rather appears in two steps (see arrow in figure 3(a)).

We now discuss the current-induced switching behaviour for positive magnetic fields. The junction was first set into the *P* state by a large positive magnetic field. Current sweeps were then systematically recorded upon lowering the magnetic field. As an example, we show a current sweep at 300 Oe in figure 3(b). Hysteretic switching can clearly be observed. Note, that the current-induced change in resistance is significantly smaller ($\Delta R/R \approx 1.9$ %) than the values obtained in the magnetic field sweep (see green dotted lines in figure 3 as guide to the eye). For comparison, we added the current-induced high resistance values obtained at 0.1 mA in figure 3(a) as red open circles. It is obvious that we cannot reach the *AP* state of the junction at any magnetic field. We rather switch into a different magnetic state of the device. This new magnetic state is stable even for current values up to 20 mA. Note that the critical field value below which we observe current-induced switching does not match the coercive field of the free layer ($H_c$ = 1200 Oe), which again indicates that we are not switching into the *AP* state. We want to emphasize that we observe such a switching into an intermediate state in all MBE grown samples independent of the junction area (ranging from $30 \times 60$ nm$^2$ to $50 \times 150$ nm$^2$) and the interlayer Cu thickness (ranging from 10 nm to 25 nm). In contrast, we have never observed the *I* state in junction, which were fabricated by sputter deposition [22]. A detailed analysis of the magnetic configuration of the intermediate state is beyond the scope of the paper and will be published elsewhere [23].

In figure 4 we summarize the threshold currents for current-induced magnetization reversal in a false-colour plot of the differential resistance as a function of both the current and the magnetic field. In order to easier visualize the switching behaviour we subtract the parabolic background in the d*V*/d*I* vs. *I* curves (see figure 3(b)). We determine a parabolic fit of the low resistance *P* state, which we extrapolate to the full current range and subtract from the measured data. We apply this method for both sweep directions (+16 mA to -16 mA and -16 mA to +16 mA) and thereafter average the differential resistance for each magnetic field. Using this method, we can easily distinguish between the low resistance *P* state (dark blue regime, figure 4) and the high resistance intermediate state *I* (green regime) as well as the hysteretic switching regime (light blue regime).

Four different regimes can be identified in the phase diagram. For large negative currents the magnetizations of both Co layers are aligned in a parallel configuration. For magnetic fields smaller than 600 Oe the thin Co layer can hysteretically be switched by the current from the parallel into the intermediate state. The critical switching currents for both the *P* to *I* and the *I* to *P* transitions merge near 600 Oe (see dotted lines in figure 4 as guide to the eye). No switching can be observed above 600 Oe. Instead, non-hysteretic peaks are observed in the differential resistance (see inset of figure 6), which indicates an unstable regime. Although we never reach the *AP* state by current sweeps in our devices, our phase diagram has striking similarities to previous results on sputtered Co | Cu | Co samples [11-12,22,24]. The main difference for our MBE grown samples is that we switch into a stable intermediate state and not into the *AP* state. Note, that in all other studies, current-induced switching is observed right below the coercive field of the free layer and the junctions can be switched into the *AP* state at all positive fields.

Although we do not reach the *AP* state by current, our phase diagram also shows an unstable regime. In comparison to previous microwave studies on the sputtered samples, it is therefore interesting to explore the magnetization dynamics of our samples in this regime. The microwave emission of the junction has been detected in the frequency domain by spectrum analysis. Figure 5(a) depicts selected

HF spectra measured at 740 Oe on the identical junction as discussed in figures 3 and 4. The spectra are corrected by a background reference spectra of the transmission line which gives signals on the order of – 60 dBm. For better illustration, a false colour plot of all spectra is plotted in figure 5(b). No peaks are observed in the spectra for low currents. As the current is increased into the unstable regime up to 10 mA, a peak appears at $f = 4.5$ GHz. When further increasing the current, the peak shifts linearly to larger frequency. The power output also increases and does not even saturate at large currents. The observed blue shift indicates an out-of-plane precession of the free layer magnetization vector [25-26]. This behaviour we only observe in a narrow field regime. A detailed analysis of the overall spectra in the phase diagram will be given elsewhere [23]. The existence of these narrow spectra furthermore indicates that the intermediate state is given by a well-defined magnetization and does not originate from a simple multi-domain configuration.

Typically, we observe the frequency blue shift in samples with a thick Cu spacer layer thickness. In contrast, samples with thinner Cu layer thickness show a frequency red shift, i.e. a decrease of spinwave frequency with increasing current. As an example, we show in figure 6 microwave spectra of a Co/Cu/Co bilayer sample with a Cu layer thickness of 15 nm and a cross section of $50 \times 100$ nm$^2$. As for the spectra in figure 5, we chose the same magnetic field strength of 740 Oe. The decrease of the spin wave frequency with increasing current can clearly be seen in the spectra. The corresponding non-hysteretic $\Delta dV/dI$ vs. $I$ curve is plotted in the inset of figure 6(a). There is a direct correlation between the emitted microwave power and the differential resistance measurement shown by the colour code in figure 6(a) with the corresponding resistance data in the inset. Consistent with previous studies the onset of the magnetization dynamics typically occurs only near the peak in the differential resistance, while the relative position of this onset varies with the magnetic field (not shown). Such a red shift in frequency is most commonly observed for in-plane magnetic fields [25]. Note, that the emitted microwave power for the junction with in-plane precession (figure 6) is an order of magnitude less than for the junction showing the out-of-plane precession (figure 5), which indicates a larger precessional angle for latter case.

## 4. Spin transfer studies in magnetic single layer systems

Most spin transfer devices consist of at least two ferromagnetic layers. One of these layers provides the spin-polarized current and at the same time may act as a fixed reference layer for the detection of the magnetization reversal of a free second ferromagnetic layer. Recently, it has been demonstrated that the distinction between a fixed and a free ferromagnetic layer is not necessary [16]. Current-induced reversible changes in the resistance have been observed in junctions with only one ferromagnetic layer for magnetic fields perpendicular to the plane of the layer. These resistance changes have been attributed to the onset of non uniform spin wave modes. Even hysteretic switching has been observed at smaller perpendicular magnetic fields. The importance of an asymmetric spin accumulation on both sides of the ferromagnetic layer - which can be realized by asymmetric leads - has been elaborated both theoretically and experimentally [13-14,16]. However, the actual magnetic microstructure in these junctions is not yet known. As crystallinity and interface roughness may affect the generation of nonuniform spin waves, we investigate the current-induced switching behaviour in MBE grown single-layer junctions at room temperature.

In figure 7 we show magneto-transport measurements on a single layer junction with a stack sequence of 5 nm Cu | 8 nm Co | 100 nm Cu and a junction area of $30 \times 60$ nm$^2$. The data were taken at room temperature with magnetic fields applied perpendicular to the thin film plane. The current sweeps show pronounced reversible dips in the differential resistance (see figure 7(a)), which may occur for both current polarities. The dips move to larger current values with decreasing magnetic field (see figure 7(b)). We observe hysteretic switching in magnetic field ranges between - 4.4 and -3 kOe and 2.8 and 4.8 kOe. The resistance changes by ~ 0.25 %. This effect cannot be explained by the anisotropic magnetoresistance (AMR) which is significantly smaller (~ 0.1%, data not shown). Similar results have been published for junctions, which had been sputtered [16-17] or had been deposited by e-beam evaporation [18].

## 5. Conclusion

In summary, we have used nanostencil mask templates to prepare magnetic nanopillars by molecular beam epitaxy. We fabricated both single layer and double layer junctions with a stacking sequence of Cu | Co | Cu and Co | Cu | Co, respectively. Double layer junctions show a magnetoresistance of up to 3 % at room temperature. By current sweeps these junctions cannot be switched from a parallel low-resistance state into the antiparallel high-resistance state. Instead, the free layer switches into a new magnetization state which results into an intermediate stable resistance of the device. This intermediate state has previously not been observed in sputtered samples. Furthermore we observed two distinct spin wave modes indicating both in-plane and out-of-plane precessions of the thin layer magnetization. For asymmetric single layer junctions our results give indirect evidence for a nonuniform magnetization distribution resulting in current-induced hysteretic switching with a magnetoresistance effect larger than expected from the AMR effect.

We acknowledge helpful discussions with B. Özyilmaz. Work was supported by the DFG through SPP 1133 and by the HGF.

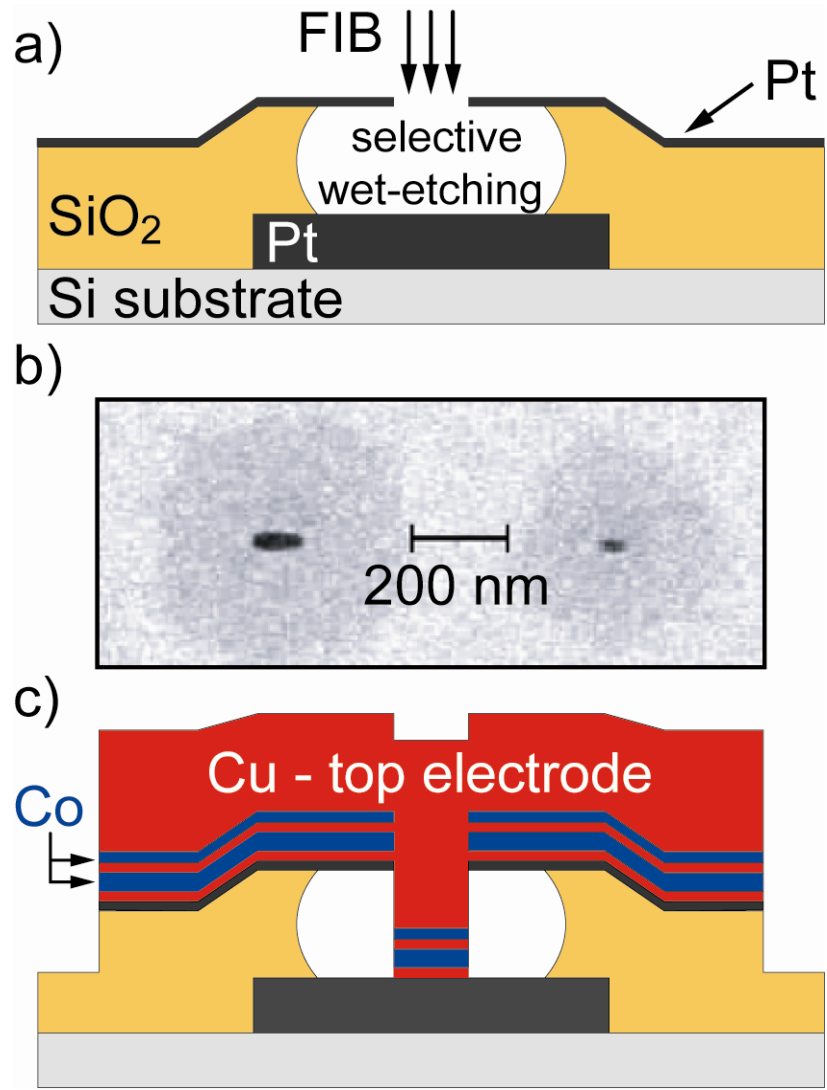

**Figure 1:** Nanostencil mask fabrication. (a) A prefabricated template, consisting of a patterned bottom electrodes (Pt) covered by an insulator (SiO$_2$) and a Pt hard mask is opened up using focused ion beam (FIB) milling. A selective wet-etching generates an undercut and gives access to the bottom electrode. (b) SEM top view of the FIB-generated holes after the selective wet-etching. (c) The desired thin film stack (here Co/Cu/Co) and Cu top electrode is grown using molecular beam epitaxy.

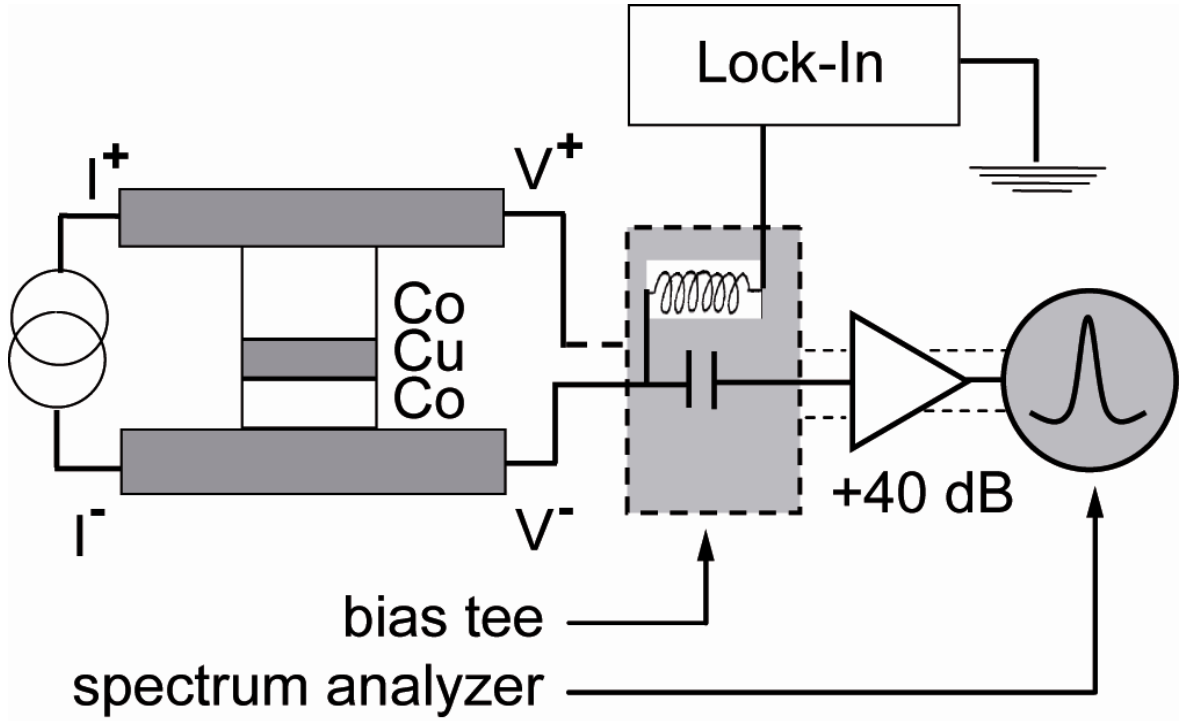

**Figure 2:** Schematic of the sample and the circuit used for differential resistance and high-frequency measurements.

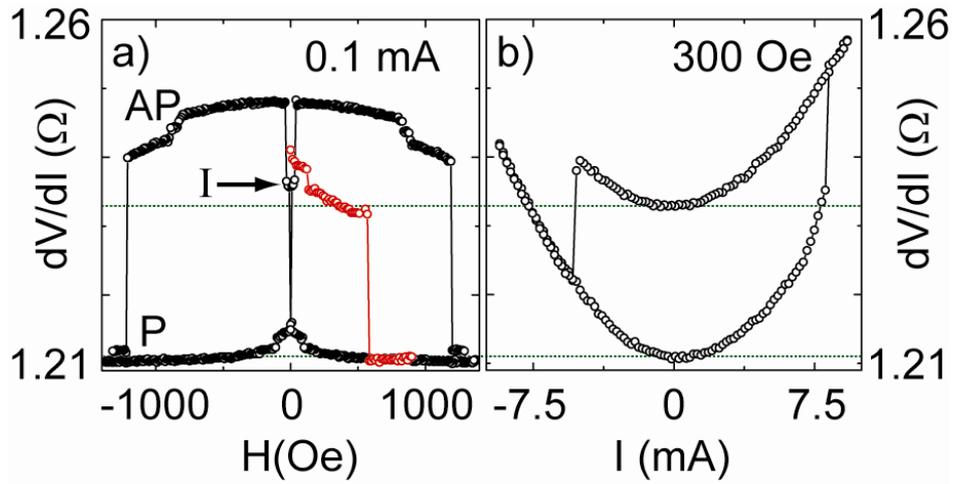

**Figure 3:** Differential resistance measurements of a magnetic double layer junction (3 nm Co | 25 nm Cu | 15 nm Co, cross-sectional area $50 \times 150$ nm$^2$) at $T = 300$ K. (a) Magnetoresistance loop with the external magnetic field applied in the sample plane along the easy axis direction (black). Red dots represent the high resistance resulting from current sweeps. (b) Current sweep at $H = 300$ Oe. The dotted green lines are guides to the eye and represent the respective low and high resistance state at 0.1 mA.

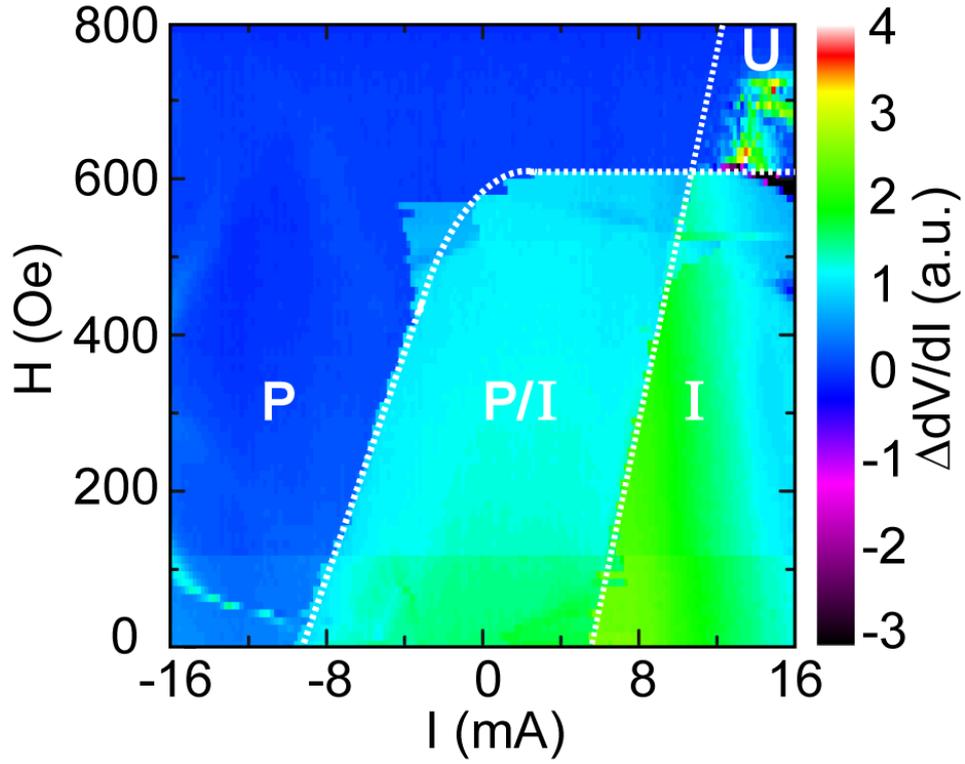

**Figure 4:** False colour plot of the differential resistance dV/dI, which is subtracted from the parabolic background (see figure 3(b)) and averaged over both current sweep directions for a magnetic double-layer junction (3 nm Co | 25 nm Cu | 15 nm Co, cross-sectional area $50 \times 150$ nm$^2$). Dashed lines indicate the boundaries between different magnetization configurations of the junction: low resistance (parallel alignment P), high resistance (intermediate state I), hysteretic regime (P/I) and unstable regime (U) for large fields and large positive currents.

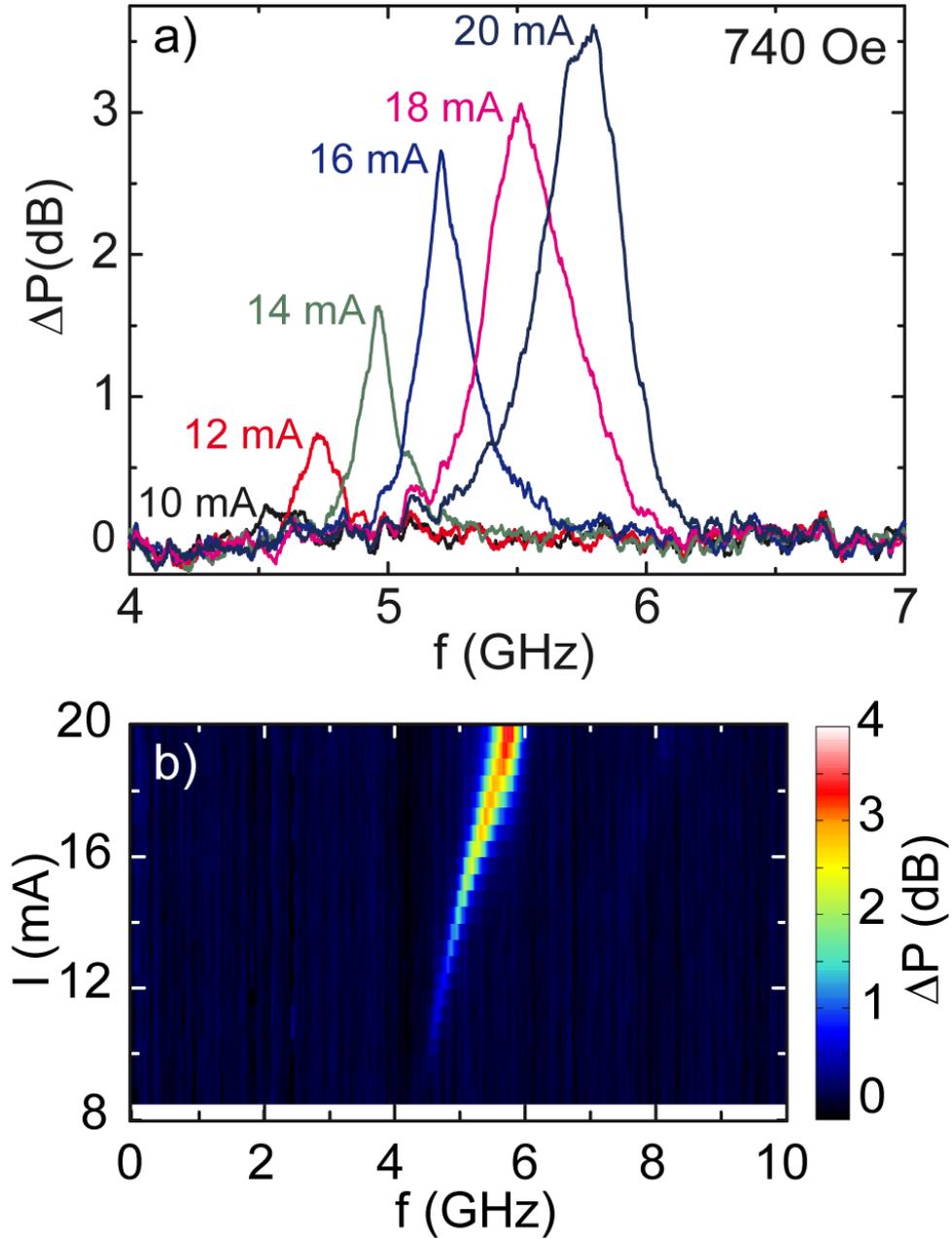

**Figure 5:** Microwave emission spectra from a magnetic double layer junction (3 nm Co | 25 nm Cu | 15 nm Co, cross-sectional area 50 × 150 nm$^2$) at $H$ = 740 Oe and $T$ = 300 K for large positive currents. (a) Selected spectra at different current values. (b) two-dimensionl false colour plot of the emitted microwave power as a function of current.

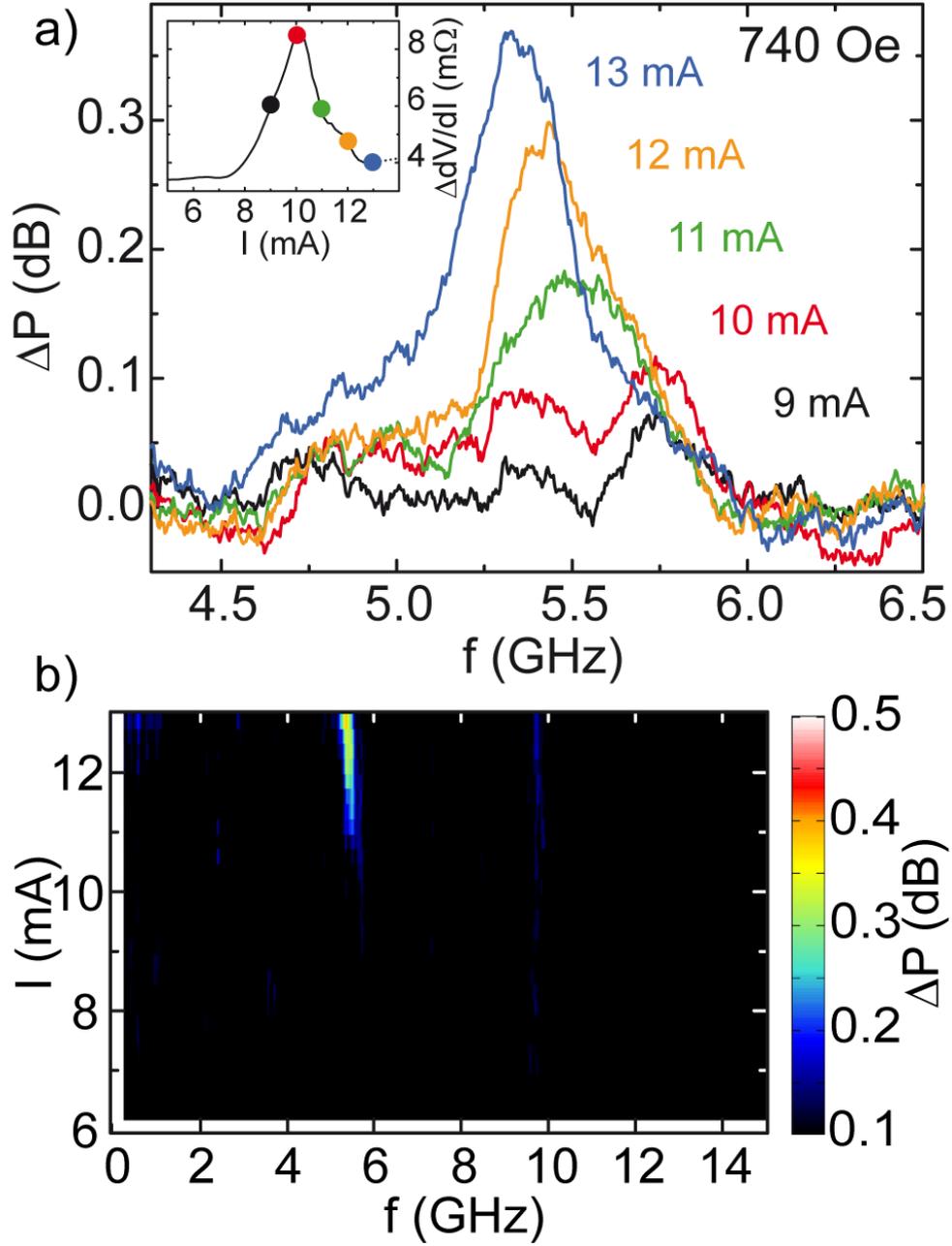

**Figure 6:** Microwave emission spectra from a magnetic double layer system (3 nm Co | 15 nm Cu | 15 nm Co; 50 × 100 nm$^2$) at $H$ = 740 Oe and $T$ = 300 K. (a) Selected spectra at different current values which are marked in the differential resistance measurement (inset). (b) two-dimensionl false colour plot of the emitted microwave power as a function of current.

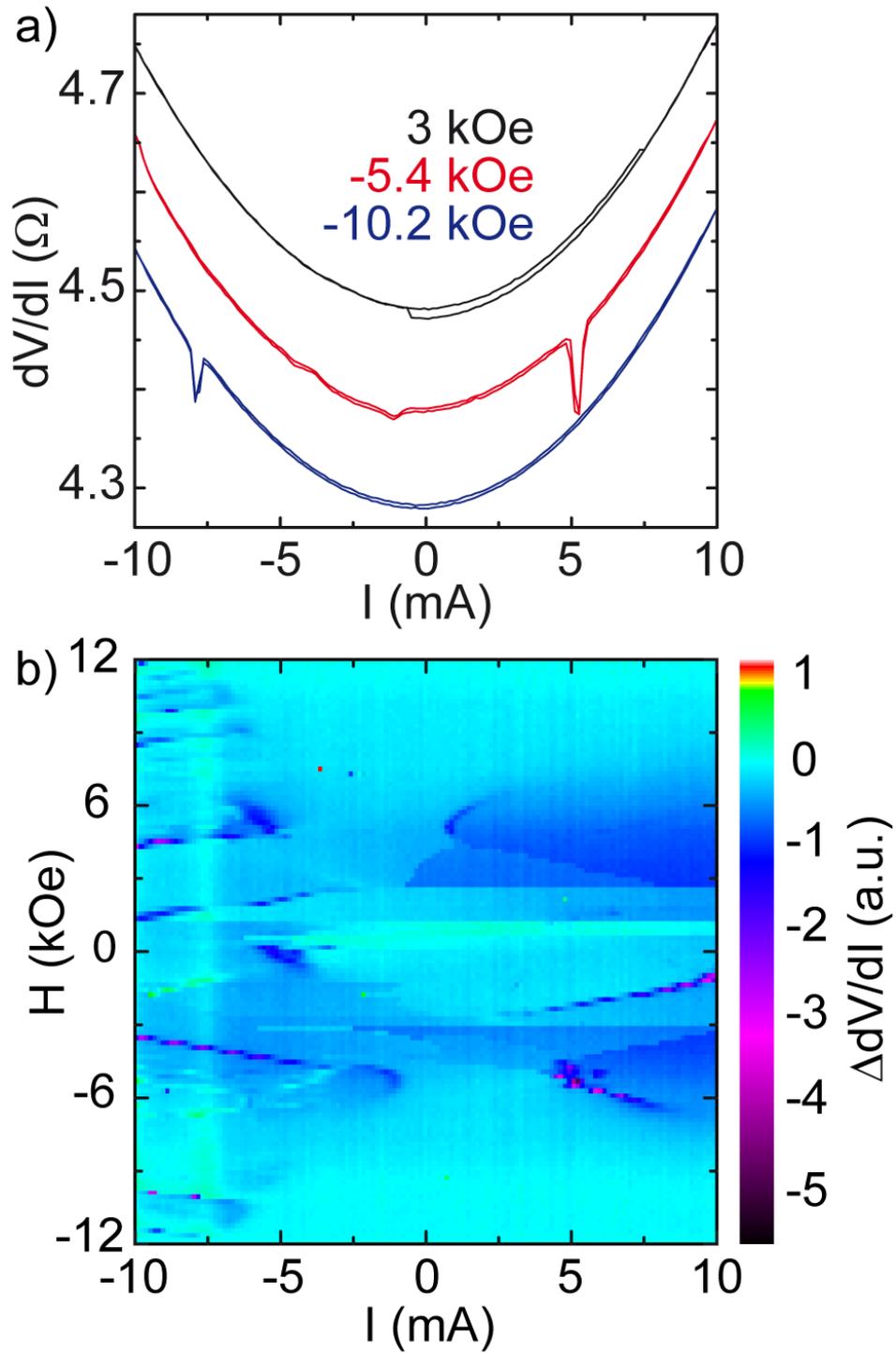

**Figure 7:** (a) Current sweeps at various magnetic fields for a single magnetic layer junction (3 nm Co | 25 nm Cu | 15 nm Co, 50 × 150 nm$^2$) with the magnetic field aligned in the out-of-plane direction. Data were taken at room temperature. (b) False colour plot of differential resistance vs. current.